\documentclass[review,3p,times]{elsarticle} 

\usepackage{setspace}

\usepackage{natbib}
 \setcitestyle{authoryear,open={(},close={)}} 

\usepackage{amssymb}
\usepackage{lipsum}
\usepackage{amsmath}
\usepackage{cool}
\usepackage{mathtools}
\usepackage{dirtytalk}
\usepackage{cuted}
\usepackage{relsize}
\usepackage[font=small]{caption}
\usepackage{subcaption}
\usepackage{ifthen}
\usepackage[normalem]{ulem}
\usepackage{xcolor}
\usepackage{url}
\usepackage[inkscapearea=page]{svg}
\usepackage{tabularx}
\usepackage{hyperref}

\usepackage{stfloats}

\usepackage{amsthm}

\usepackage{tikz}

\usepackage{lineno}

\usepackage{fdsymbol}

\begin{document}

\begin{frontmatter}

\title{Level set-based inverse homogenisation of three-dimensional piezoelectric materials}

\author{Zachary J. Wegert\corref{cor3}}
\ead{zach.wegert@hdr.qut.edu.au}
\author{Anthony P. Roberts}
\ead{ap.roberts@qut.edu.au}
\author{Vivien J. Challis\corref{cor3}}
\ead{vivien.challis@qut.edu.au}
\cortext[cor3]{Corresponding author}
\address{School of Mathematical Sciences, Queensland University of Technology, Brisbane, QLD 4000, Australia}

\begin{abstract}
In this paper we use memory-distributed level set-based topology optimisation to design three-dimensional periodic piezoelectric materials with enhanced properties. We compare and assess several existing iterative solvers with respect to their weak scalability and find that an approximate Schur complement preconditioned generalized minimal residual method method demonstrates the best performance and scalability for solving the piezoelectric homogenisation equations. We use the developed techniques to computationally design high-resolution piezoelectric metamaterials with enhanced stiffness and piezoelectric properties that yield new insights into material design for sensor, hydrophone, and actuator applications. We suggest two robust structures with no fine-scale features that exhibit enhanced piezoelectric properties several times larger than those of the base material. We find that level set-based topology optimisation is well suited to problems involving piezoelectricity and has the advantage of avoiding large regions of intermediate density material. Our memory-distributed level-set implementation is open source and provided for practitioners in the community.
\end{abstract}



\end{frontmatter}


\section{Introduction}
\label{sec:Introduction}
A piezoelectric material exhibits the phenomenon in which deformation generates electric charge, and vice versa \citep{IEEE}. These materials are used in several engineering applications including sensing, actuator technology, and energy harvesting \citep{AsifKhan2016,Chopra2002,Sodano2004}. 

Recent advancements in fine-scale additive manufacturing techniques \cite[e.g.,][]{10.1002/adfm.202005141_2020,10.1088/1757-899X/455/1/012046_2018,10.1038/s41563-018-0268-1_2019} have prompted the design of state-of-the-art heterogeneous piezoelectric metamaterials that exhibit exotic macroscopic properties \citep{doi:10.1177/1464420721995858,Yang2019Designingmetamaterial}. These have been shown to significantly improve material properties and figures of merit for engineering applications. \citet{Challagulla2012} showed that network structures are able to produce a wide range of piezoelectric properties. \citet{Yang2019Designingmetamaterial} proposed novel metamaterials with full non-zero piezoelectric charge constants. It is hypothesised that these types of materials could be used to design multi-functional devices such as sensors or energy harvesters. \citet{Khan2019} and \citet{Khan2019-33} proposed auxetic network metamaterials that significantly improved the figures of merit related to hydrophone applications. \citet{Cui2019} proposed the design and additive manufacture of a wide variety of three-dimensional piezoelectric metamaterials based on network structures. They further showed that the metamaterials could be utilised for impact detection and force directionality sensing owing to their tailorable and high-performance properties. \citet{Shi2019} accounted for both thermal properties and piezoelectric properties in periodic metamaterials with different pore shapes. They found that the effective pyroelectric and piezoelectric material properties could be significantly improved by adjusting the pore structure. \citet{XU2020108417} showed that triply periodic minimal surfaces could be used to enhance piezoelectric material properties. \citet{Wegert_2021} demonstrated computationally that multi-poled piezoelectric materials can possess full non-zero piezoelectric coefficients and provide excellent piezoelectric properties for sensing applications.

Although density-based topology optimisation techniques have been applied to optimisation of macroscopic piezoelectric devices
\citep[e.g.,][]{KIM20103153,Zhang_Kang_2014,Zhang_Takezawa_Kang_2018,Moretti_Silva_2019,Guzmán_Silva_Rubio_2020,He_2021,Homayouni_2021,Yang_Cheng_Wang_Meng_Homayouni-Amlashi_2022,Wegert_Roberts_Bandyopadhyay_Challis_2023,Homayouni_et_al2024}, only a few works focus on microstructure optimisation. \cite{10.1007/s004660050188_1997,10.1016/S0045-7825(98)80103-5_1998} considered optimising piezoelectric figures of merit subject to constraints on the stiffness tensor using a density method and linear programming. \cite{Wegert_2022} used multi-objective density-based microstructure optimisation to design periodic piezoelectric metamaterials with high-performance piezoelectric properties for hydrophone and sensor applications. The authors found that intermediate densities increase the objective in a non-physical manner. As a result, the optimiser tends to favour intermediate material in areas that can artificially improve the objective. This phenomena has been previously observed by \cite{KIM20103153} and they proposed more appropriate choices of penalisation exponents in the context of Solid Isotropic Material with Penalisation (SIMP). However, appropriate exponents do not completely eliminate the issue \citep{Wegert_2022}. To circumvent this problem, \cite{Fang_Meng_Zhou_Wang_Guo_2023} used the moving morphable void method to design of piezoelectric actuators. In addition, \cite{Stankiewicz_Dev_Weichelt_Fey_Steinmann_2024} recently proposed a sequential hybrid method for 2D piezoelectric microstructure optimisation that first solves a density-based topology optimisation problem then a shape optimisation problem. In this work, we utilise \textit{level set-based topology optimisation} \citep{10.1016/bs.hna.2020.10.004_978-0-444-64305-6_2021}, bypassing the aforementioned problem that occurs with density-based topology optimisation because conventional level-set methods are free of large areas of intermediate densities and instead smoothly interpolate material properties near the zero level-set contour \citep{DijkLevelSetReview}. To the best of our knowledge, level set-based methods have not been previously utilised for piezoelectric microstructure optimisation.

Of all the works discussed above, only a few consider piezoelectric optimisation problems in three dimensions \citep{10.1016/S0045-7825(98)80103-5_1998,Wegert_2022,Wegert_Roberts_Bandyopadhyay_Challis_2023,Homayouni_et_al2024}. Moreover, these works considered low mesh resolutions to make the optimisation problems computationally tractable in serial implementations. In this work, we implement three-dimensional level-set topology optimisation in parallel using the programming language Julia \citep{doi:10.1137/141000671} and the package GridapTopOpt \citep{GridapTopOpt}. This package relies on the Gridap finite element ecosystem \citep{Badia2020} and provides an extendable framework for assembling and solving PDE-constrained optimisation problems that can be readily distributed across several cores of a personal computer or high-performance computing cluster. For the PDEs describing piezoelectricity, the numerical discretisation has a block structure that is either symmetric but indefinite or positive definite but nonsymmetric \citep{Cao_Neytcheva_2021}. As a result, standard iterative methods take hundreds or even thousands of iterations to converge and are not scalable. In this work we compare a na\"ive diagonal block preconditioned generalized minimal residual method (DBP-GMRES), tridiagonal conjugate gradient (TriCG) \citep{Montoison_Orban_2021}, and an approximate Schur complement preconditioned GMRES method (SCP-GMRES) \citep{Saad_2003,Cao_Neytcheva_2021}. These developments allow us to push the computational resolution far beyond anything previously achieved in the literature. 

In this paper we consider two distinct types of optimisation problems. The first maximises the effective bulk modulus subject to a constraint on the volume and the effective hydrostatic strain coefficient. The bulk modulus is a measure of stiffness while the hydrostatic strain coefficient is a measure of effectiveness of the electromechanical coupling \citep{Challagulla2012, 10.1007/s004660050188_1997}. The output of this optimisation problem is a multi-functional metamaterial that provides enhanced stiffness for a set electromechanical coupling capability. The optimisation problem bears similarity to the one considered by \cite{Wegert_2022}, where the problem was instead posed as maximising a linear combination of bulk modulus and hydrostatic strain coefficient. We consider this problem using the effective Voigt bulk modulus as our measure of the effective bulk modulus, for which a uniform strain is assumed \citep{Hill_1952}. We also consider maximising the effective Reuss bulk modulus subject to the same constraints. The Reuss bulk modulus assumes a uniform stress and has been shown to always be less than or equal to the Voigt bulk modulus with equality for isotropic materials \citep{Hill_1952}. For piezoelectric materials in which the underlying stiffness tensor is not isotropic, the Reuss and Voigt bulk moduli are different. For this reason, we compare both objectives in this paper. The second type of optimisation problem that we consider is maximising the effective hydrostatic strain coefficient subject to a constraint on the effective stiffness in the poling direction. The output of this optimisation problem is a metamaterial that exhibits extremal electromechanical coupling. It is similar to the optimisation problem considered by \cite{10.1016/S0045-7825(98)80103-5_1998}. By considering both types of optimisation problem at high computational resolutions with the level-set method, we are able to gain new insights into the design of piezoelectric metamaterials for sensing and hydrophone applications.  

The remainder of the paper is as follows. In Section \ref{sec:Numerical methods} we discuss the numerical methods including an analysis of solvers for the piezoelectric state equations. In Section \ref{sec:Results and discussion} we present and discuss the optimisation results. Finally, in Section \ref{sec: Conclusions} we present our concluding remarks.

\section{Numerical methods}
\label{sec:Numerical methods}

\subsection{The level-set method}
Level-set methods implicitly track a domain $\Omega$ inside a bounding domain $D\subset\mathbb{R}^d$ using the zero level set of a scalar function $\phi:D\rightarrow\mathbb{R}$ \citep{978-0-521-57202-6_1996,978-0-387-22746-7_2006}. We follow the classical approach and define the level-set function to satisfy
\begin{equation}
\begin{cases}\phi(\boldsymbol{x})<0&\text{if } \boldsymbol{x} \in \Omega, \\ \phi(\boldsymbol{x})=0 &\text{if } \boldsymbol{x} \in \partial \Omega, \\\phi(\boldsymbol{x})>0 &\text{if } \boldsymbol{x} \in D \backslash \bar{\Omega}.\end{cases}
\end{equation}
The evolution of the boundary $\partial\Omega$ for small time $t\in(0,T)$ under a normal velocity field $v$ is described by the Hamilton-Jacobi evolution equation \citep{978-0-521-57202-6_1996,978-0-387-22746-7_2006,10.1016/bs.hna.2020.10.004_978-0-444-64305-6_2021}:
\begin{align}\label{eqn: HJ}
    \begin{cases}
    \pderiv{\phi}{t}(t,\boldsymbol{x}) + v(\boldsymbol{x})\lvert\nabla\phi(t,\boldsymbol{x})\rvert = 0,\\
    \phi(0,\boldsymbol{x})=\phi_0(\boldsymbol{x}),\\
    \boldsymbol{x}\in D,~t\in(0,T),
    \end{cases}
\end{align}
where $\phi_0(\boldsymbol{x})$ is the initial condition for $\phi$ at $t=0$. To optimise a functional $J(\Omega)$, the normal velocity $v$ in Eq.~\eqref{eqn: HJ} is inferred by using the first-order change of $J$ under a change in $\Omega$. This can be understood using a shape derivative \citep{10.1016/j.jcp.2003.09.032_2004,10.1016/bs.hna.2020.10.004_978-0-444-64305-6_2021}. 

We refer the reader to our recent work \citep{GridapTopOpt} for details of our level-set topology optimisation approach and relevant mathematical background and references.

\subsection{Linear piezoelectric homogenisation}
The state equations for linear pieozelectric homogenisation over a domain $\Omega$ contained in a representative volume element $D\subset\mathbb{R}^3$ under an applied strain field $\bar{\varepsilon}_{i j}$ or an applied electric field $\bar{E}_{i}$ are \citep{YvonnetCompHomogenization}
\begin{align}
    -\sigma_{ij,i}&= 0\text{ in }\Omega, \label{eqn:sigma_ij,i_PZ}\\
    D_{i,i}&= 0\text{ in }\Omega, \\
    \sigma_{ij}n_j&= 0~\text{on }\partial\Omega,\\
    D_{i}n_i&= 0~\text{on }\partial\Omega,\\
    \frac{1}{\operatorname{Vol}(\Omega)}\int_{\Omega}{\varepsilon_{kl}}~\mathrm{d}\boldsymbol{x}&=\bar{\varepsilon}_{kl},\\
    \frac{1}{\operatorname{Vol}(\Omega)}\int_{\Omega}{E_{i}}~\mathrm{d}\boldsymbol{x}&=\bar{E}_{i},\label{eqn:pz_state_last_eqn}
\end{align}
with
\begin{align}
    \sigma_{i j}&=C^E_{i j k l} \varepsilon_{k l}-e_{k i j} E_{k},  \\
    D_{i}&=e_{i j k} \varepsilon_{j k}+\kappa^\varepsilon_{i k} E_{k}, \\
    \varepsilon_{ij}&=\frac{1}{2}\left(u_{i,j}+u_{j,i}\right), \\
    E_i &= -\varphi_{,i}.
\end{align}
In the above $\sigma_{ij}$ is the stress tensor, $D_i$ is the electric displacement vector, $E_i=E_i(\varphi)$ is the $\Omega$-periodic electric field with electric potential $\varphi$, $\varepsilon_{ij}=\varepsilon_{ij}(\boldsymbol{u})$ is the $\Omega$-periodic strain field with displacement $\boldsymbol{u}$, $C^E_{i j k l}$ is the spatially dependent elasticity tensor at a fixed electric field, $e_{ijk}$ is the spatially dependent piezoelectric tensor, and $\kappa^\varepsilon_{ij}$ is the spatially dependent dielectric tensor at a fixed strain.

To compute the homogenised material tensors of a periodic piezoelectric material, the above state equations are solved over $\Omega$ for nine different combinations of macroscopic strain fields and electric fields. Macroscopic strains $\bar{\varepsilon}_{i j}$ and electric fields $\bar{E}_i$ are applied by decomposing the fields as $\varepsilon_{i j}=\bar{\varepsilon}_{i j}+\tilde{\varepsilon}_{i j}$ and $E_{i}=\bar{E}_{i}+\tilde{E}_{i}$, respectively. In these expressions, $\tilde{\cdot}$ denotes the fluctuation field. The macroscopic strain fields are then given by the unique components of $\bar{\varepsilon}_{i j}^{(k l)}=\frac{1}{2}\left(\delta_{i k} \delta_{j l}+\delta_{i l} \delta_{j k}\right)$ in $k$ and $l$, while the macroscopic electric fields are given by the unique components of $\bar{E}_i^{(j)}=\delta_{ij}$ in $j$. The notation $\tilde{\varepsilon}_{i j}^{(k l)}$ denotes the strain field fluctuation arising from the applied strain field $\bar{\varepsilon}_{i j}^{(k l)}$, while $\tilde{E}_{i}^{(j)}$ denotes the electric field fluctuation arising from the applied electric field $\bar{E}_{i}^{(j)}$. Using these decompositions, Eqs.~\eqref{eqn:sigma_ij,i_PZ}-\eqref{eqn:pz_state_last_eqn} can be solved for each unique applied macroscopic field using a finite element method. The weak formulations for these problems are provided in the supplementary material.

It is important to note that owing to the vast difference in magnitudes between the stiffness tensor $C_{ijkl}^E$ (${\sim}10^{11}$), piezoelectric coupling tensor $e_{ijk}$ (${\sim}10^{1}$), and dielectric tensor $\kappa_{ij}$ (${\sim}10^{-9}$), non-dimensionalisation of the above is important to avoid poor conditioning. In this work we utilise the non-dimensionalisation proposed by \cite{Cao_Neytcheva_2021}.

Once solutions to the above have been calculated, the homogenised material coefficients can then be computed via \citep{YvonnetCompHomogenization} 
\begin{align}\label{eqn:hom_C_ijkl_ALT}
    \bar{C}_{ijkl}^E&=\frac{1}{\operatorname{Vol}(D)}\int_{\Omega}{C_{rsmp}^E(\tilde{\varepsilon}_{mp}^{(kl)}+\bar{\varepsilon}_{mp}^{(kl)})\bar{\varepsilon}_{rs}^{(ij)}-e_{mrs}\tilde{E}_m^{(kl)}\bar{\varepsilon}_{rs}^{(ij)}}~\mathrm{d}\boldsymbol{x},\\
\label{eqn:hom_e_ijk_ALT}
    \bar{e}_{ijk}&=\frac{1}{\operatorname{Vol}(D)}\int_{\Omega}{e_{rlm}(\tilde{\varepsilon}_{lm}^{(jk)}+\bar{\varepsilon}_{lm}^{(jk)})\bar{E}_{r}^{(i)}+\kappa^\varepsilon_{rl}\tilde{E}_l^{(jk)}\bar{E}_{r}^{(i)}}~\mathrm{d}\boldsymbol{x},\\
\label{eqn:hom_kappa_ij_ALT}
    \bar{\kappa}^\varepsilon_{ij}&=\frac{1}{\operatorname{Vol}(D)}\int_{\Omega}{e_{rlm}\tilde{\varepsilon}_{lm}^{(j)}\bar{E}_r^{(i)}+\kappa^\varepsilon_{rl}(\tilde{E}_l^{(j)}+\bar{E}_l^{(j)})\bar{E}_r^{(i)}}~\mathrm{d}\boldsymbol{x}.
\end{align}
For the reminder of this work, we assume that $D$ is a unit cell with $\operatorname{Vol}(D)=1$. The shape derivatives of these homogenised quantities are provided in the supplementary material.

\subsection{Optimisation problem}
In this paper we consider two types of optimisation problems. The first maximises the effective Voigt or Reuss bulk modulus ($\bar{B}_{\mathrm{Voigt}}$ and $\bar{B}_{\mathrm{Reuss}}$ respectively) subject to constraints on the volume $\operatorname{Vol}(\Omega)$ and the effective hydrostatic strain coefficient $\bar{d}_h(\Omega)$. We choose to formulate this optimisation problem using an equality constraint on the volume so that we can explore the bulk modulus and hydrostatic coupling cross-property space for a specified volume fraction. The hydrostatic strain coefficient is a measure of effectiveness of the electromechanical coupling \citep{Challagulla2012, 10.1007/s004660050188_1997}. The calculation of the Voigt bulk modulus assumes that a uniform strain is applied, while the Reuss bulk modulus assumes a uniform stress \citep{Hill_1952}. It has been shown that $\bar{B}_{\mathrm{Reuss}} \leq\bar{B}_{\mathrm{Voigt}}$ with equality for isotropic materials \citep{Hill_1952}. For the case of piezoelectric materials, in which the underlying stiffness tensor is not isotropic, the Reuss and Voigt bulk moduli are different. For this reason we compare both objectives here. The optimisation problem for maximising $\bar{B}_\mathrm{Voigt}(\Omega)$ is given by
\begin{equation}\label{eqn:optim prob I}
          \begin{aligned}
            \underset{\Omega\subset D}{\max}~
              &\bar{B}_\mathrm{Voigt}(\Omega)\\
            \text{s.t.}~~ & \operatorname{Vol}(\Omega)=V^*,\\
            &\bar{d}_h(\Omega)=d_h^*,\\
            &a(\boldsymbol{u},\boldsymbol{v})=l(\boldsymbol{v}),~\forall \boldsymbol{v}\in V.
          \end{aligned}
\end{equation}
From here on, we omit the dependence on $\Omega$ to help the clarity of notation. In the above, the final line denotes the weak formulation of the piezoelectricity homogenisation equations, $\bar{B}_\mathrm{Voigt}$ is given by $\bar{B}_\mathrm{Voigt}=\frac{1}{9}\bar{C}^E_{iijj}$, and $\bar{d}_h$ is given by $\bar{d}_h=\bar{d}_{zxx}+\bar{d}_{zyy}+\bar{d}_{zzz}$ 
with $\bar{d}_{ijk}=\bar{e}_{ilm}\bar{S}^E_{lmjk}$ and the effective compliance tensor $\bar{S}^E_{ijkl}$ is the inverse of the effective stiffness tensor $\bar{C}^E_{ijkl}$. The latter inverse can be computed in Voigt notation \citep{IEEE}. The optimisation problem for maximising $\bar{B}_\mathrm{Reuss}$ is analogous to Eq.~\eqref{eqn:optim prob I} where the optimisation functional is replaced with $\bar{B}_\mathrm{Reuss}=1/\bar{S}^E_{iijj}$. The shape derivatives of the quantities above can be calculated exactly using the results in the supplementary material. The problem posed in Eq.~\eqref{eqn:optim prob I} is a reformulation of the one considered by \cite{Wegert_2022}, where the objective was instead posed as a linear combination of $\bar{B}_h$ and $\bar{d}_h$ with a constraint on the volume.

The second optimisation problem that we consider is maximising the effective hydrostatic strain coefficient $\bar{d}_h$ subject to a constraint on the effective stiffness $\bar{C}_{zzzz}$ in the piezoelectric poling direction. Roughly speaking, the poling direction is the direction in which an electric field is generated under axial deformations. This optimisation problem can be written as
\begin{equation}\label{eqn:optim prob 2}
          \begin{aligned}
            \underset{\Omega\subset D}{\max}~
              &\bar{d}_h(\Omega)\\
            \text{s.t.}~~ & \operatorname{Vol}(\Omega)=V^*,\\
            &\bar{C}_{zzzz}(\Omega)=C_{zzzz}^*,\\
            &a(\boldsymbol{u},\boldsymbol{v})=l(\boldsymbol{v}),~\forall \boldsymbol{v}\in V.
          \end{aligned}
\end{equation}
This optimisation problem seeks to find optimised metamaterials that exhibit extremal electromechanical coupling and is similar to one considered by \cite{10.1016/S0045-7825(98)80103-5_1998}. An equality constraint on the volume is used for consistency with the first optimisation problem.

The base material used for all computations is chosen to be $z$-poled PZT-5A with material constants \citep{DesignOfPiezocompositePart1}
\begin{align}
    &C^{E}_{pq}=\begin{pmatrix*}[r]
    12.04 & 7.52 & 7.51 & 0.0&0.0&0.0\\
    7.52 & 12.04 & 7.51 & 0.0&0.0&0.0\\
    7.51&  7.51 & 11.09 & 0.0&0.0&0.0\\
    0.0&0.0&0.0& 2.1 & 0.0&0.0\\
    0.0&0.0&0.0& 0.0&2.1&  0.0\\
    0.0&0.0&0.0& 0.0&0.0&2.3
    \end{pmatrix*}\times10^{10},\\
&e_{ip}=\begin{pmatrix*}[r]
0.0     &  0.0  &     0.0&0.0     &  12.3 &  0.0\\
    0.0   &    0.0   &    0.0&12.3 & 0.0    &   0.0\\
   -5.4 &  -5.4  & 15.8  & 0.0     &  0.0    &   0.0
    \end{pmatrix*},\label{eqn:pz-cof-tensor}\\
&\kappa^{\varepsilon}_{ij}=\begin{pmatrix*}[r]
4.78   &  0.0&0.0\\
       0.0   &   4.78  &  0.0\\
       0.0& 0.0  &  7.35
    \end{pmatrix*}\times 10^{-9},
\end{align}%
where both $p$ and $q$ are indices in Voigt notation, and the units of $C^{E}_{pq}$, $e_{ip}$, and $\kappa^{\varepsilon}_{ij}$ are $\textnormal{N/m}^2$, $\textnormal{C/m}^2$, and $\textnormal{F/m}$, respectively. Note that the value of the hydrostatic piezoelectric strain coefficient $d_h$ for the base material is 31.7 (pC/N). 

\subsection{Memory-distributed implementation}
We implement the aforementioned optimisation problems in the Julia programming language \citep{doi:10.1137/141000671} using the GridapTopOpt package \citep{GridapTopOpt}. The source code is readily available at \url{https://github.com/zjwegert/Wegert_et_al_2024b}. GridapTopOpt provides an extendable framework for solving PDE-constrained optimisation problems by partitioning a domain and distributing it and computational processing across several distinct processors. Construction of these subdomains and all resulting computation is then done locally on each processor with communication using message passing interface (MPI) where needed. 
\begin{figure}[t]
    \centering
    \includegraphics[width=0.5\linewidth]{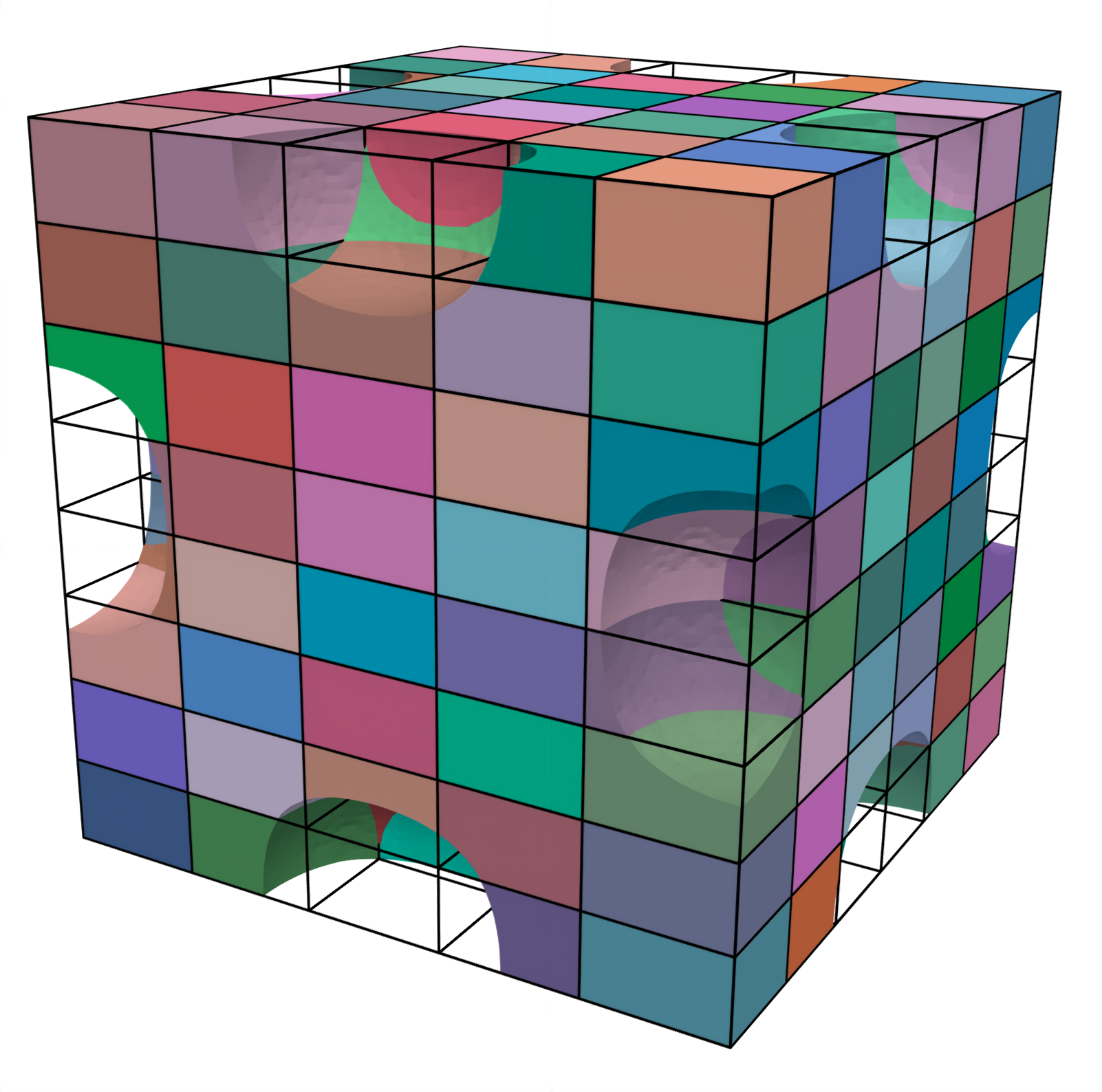}
    \caption{Visualisation of an initial structure $\Omega$ defined on a (5,6,8)-partitioned computational domain $D$ to be analysed with 240 processors. The underlying mesh has $100^3$ first-order hexahedral elements. The initial signed-distance function for $\Omega$ is generated by reinitialising the level-set function $\phi(\boldsymbol{x})=-\cos(2\pi x)\cos(2\pi y)\cos(2\pi z)-0.2$. We visualise the resulting structure using an isovolume for $\phi\leq0$.}
    \label{fig:partitioned_domain}
\end{figure}%
Figure \ref{fig:partitioned_domain} shows a visualisation of a domain that has been defined on a (5,6,8)-partitioned computational domain with 240 processors. GridapTopOpt relies on Gridap \citep{Badia2020,Verdugo2022} for serial finite element assembly from arbitrary weak formulations of PDEs; GridapDistributed \citep{Badia2022} for memory-distributed assembly; and PartitionedArrays \citep{PartitionedArrays_Verdugo2021} to handle memory-distributed linear algebra without having to resort to the low-level MPI interface provided by Julia \citep{Byrne2021}. We refer the reader to \cite{GridapTopOpt} and references therein for further details regarding GridapTopOpt.

\subsection{Linear solvers and preconditioning}
The finite element discretisation of the piezoelectric homogenisation equations takes the following block form:
\begin{equation}\label{eqn: sym-piezo}
    \underbrace{\begin{pmatrix}
        A&B^\intercal\\
        B&-C
    \end{pmatrix}}_{\mathcal{A}}\begin{pmatrix}
        \boldsymbol{x}\\
        \boldsymbol{y}
    \end{pmatrix} = \begin{pmatrix}
        \boldsymbol{a}\\
        \boldsymbol{b}
    \end{pmatrix}, 
\end{equation}
or equivalently
\begin{equation}
    \underbrace{\begin{pmatrix}
        A&B^\intercal\\
        -B&C
    \end{pmatrix}}_{\hat{\mathcal{A}}}\begin{pmatrix}
        \boldsymbol{x}\\
        \boldsymbol{y}
    \end{pmatrix} = \begin{pmatrix}
        \boldsymbol{a}\\
        -\boldsymbol{b}
    \end{pmatrix},
\end{equation}
where $A\in\mathbb{R}^{n\times n}$ and $C\in\mathbb{R}^{m\times m}$ are positive-definite symmetric matrices that correspond to the elasticity and conductivity bilinear forms in the weak formulation (see supplementary material), and $B\in\mathbb{R}^{m\times n}$ corresponds to the piezoelectric coupling term. Standard solvers have difficulty solving these types of systems because $\mathcal{A}\in\mathbb{R}^{(n+m)\times(n+m)}$ is indefinite while $\hat{\mathcal{A}}\in\mathbb{R}^{(n+m)\times(n+m)}$ is nonsymmetric. It is therefore imperative that focus be given to the development of scalable linear solvers for this type of system when developing distributed computational methods.

In this work, we compare a na\"ive diagonal block preconditioned generalized minimal residual method (DBP-GMRES), tridiagonal conjugate gradient (TriCG) \citep{Montoison_Orban_2021}, and an approximate Schur complement preconditioned GMRES method (SCP-GMRES). The latter is based on the approximate Schur complement discussed by \cite{Cao_Neytcheva_2021} and \cite{Saad_2003}. All three of these methods are readily implemented in parallel using the packages mentioned above. In the following, we briefly describe each method and compare their weak scalability.

\subsubsection{DBP-GMRES}
DBP-GMRES utilises a simple diagonal block preconditioner $\mathcal{P}_{\mathrm{DBP}}$ applied to $\mathcal{A}$ given by
\begin{equation}
    \mathcal{P}_{\mathrm{DBP}} = \begin{pmatrix}
        A&0\\
        0&-C
    \end{pmatrix}.
\end{equation}
The parallel implementation of GMRES from GridapSolvers \citep{GridapSolvers} is then used to solve the resulting right-preconditioned system. This allows us to leverage the block structure of the matrices.

Note that we require efficient solvers to evaluate the maps $\boldsymbol{v}\mapsto A\backslash \boldsymbol{v}$ and $\boldsymbol{u}\mapsto -C\backslash \boldsymbol{u}$. For these maps and the corresponding ones below, we utilise the preconditioned conjugate gradient method with algebraic multi-grid preconditioning (CG-AMG) with a relative tolerance of $\varepsilon_\mathrm{rel}$ to be later specified. These are implemented using PETSc \citep{petsc-web-page} via GridapPETSc \citep{GridapPETSc_Verdugo2021}.

\subsubsection{TriCG}
TriCG solves Eq.~\eqref{eqn: sym-piezo} using orthogonal tridiagonalisation and naturally induces an $LDL^\intercal$ factorisation. We implement the method in parallel using PartitionedArrays. This is based on the implementation in Krylov.jl \citep{montoison-orban-2023} and Algorithm 3 described by \cite{Montoison_Orban_2021}. We refer the reader to \cite{Montoison_Orban_2021} for further discussion.

Note that the method also requires maps $\boldsymbol{v}\mapsto A\backslash \boldsymbol{v}$ and $\boldsymbol{u}\mapsto -C\backslash \boldsymbol{u}$. 

\subsubsection{SCP-GMRES}
The exact Schur complement preconditioner for $\hat{\mathcal{A}}$ is typically given in lower-triangular block matrix form as
\begin{equation}
    \mathcal{P}_{\mathrm{SCP}} = \begin{pmatrix}
        A&0\\
        -B&S
    \end{pmatrix}
\end{equation}
where $S=C + B^\intercal A^{-1}B$ is called the exact Schur complement \citep{Saad_2003}. It is typically dense and requires the full inverse of $A$. As a result, approximations of the Schur complement are of interest. One such method of approximation is via a local element/cell-wise construction \citep{Saad_2003,Cao_Neytcheva_2021}. \cite{Saad_2003} discusses a general local elemental construction in terms of the finite element method, while more recently \cite{Cao_Neytcheva_2021} considered a cell-by-cell construction in the context of piezoelectricity via a radial point interpolation meshfree method. In the context of the finite element method, the approximate Schur complement can be constructed as
\begin{equation}
    \bar{S}=\sum_{\Omega_e\in\Omega}C_e + B_e^\intercal A_e^{-1}B_e,
\end{equation}
where $A_e$, $B_e$, and $C_e$ correspond to the local stiffness matrices in $A$, $B$, and $C$ respectively, and the summation denotes the usual finite element assembly process. Note that $A_e$ is typically singular, this can be remedied by adding a small perturbation to the diagonal \citep{Cao_Neytcheva_2021}. In our experience the solver is insensitive to this parameter and it can be taken to be $10^{-10}$. \cite{Cao_Neytcheva_2021} proved that the resulting approximate Schur complement $\bar{S}$ is spectrally equivalent to the exact Schur complement $S$.

GMRES is then used to solve the resulting right-preconditioned system. Note that we require the maps $\boldsymbol{v}\mapsto A\backslash \boldsymbol{v}$ and $\boldsymbol{u}\mapsto S\backslash \boldsymbol{u}$ to invert $\mathcal{P}_\mathrm{SCP}$ as part of this procedure. 

\subsubsection{Comparison}
In this section, we consider a weak scaling comparison of the above methods where the number of degrees of freedom \textit{per processor} is fixed, while the number of processors is increased. This illustrates the scaling capability of the solvers as the problem size is increased. Our benchmark considers solving the linear system resulting from the equations for piezoelectric homogenisation for the unique macroscopic strain field $\bar{\varepsilon}_{ij}^{11}$ over the domain $\Omega$ shown in Figure \ref{fig:partitioned_domain}. This domain is defined by $\phi<0$ for $\phi(\boldsymbol{x})=-\prod_{i=1}^{3}\cos(2\pi x_i)-0.2$. For this test, we fix the degrees of freedom per processor at 32,000 and consider the maximum time taken across all processors to update the matrix and preconditioner, and solve the resulting system. We then take the minimum time across 10 independent runs to find the best obtainable performance. We benchmark using Intel® Xeon® Platinum 8274 Processors with 4GB of RAM per core on the Gadi@NCI Australian supercomputer. In our testing, we consider 48, 240, 432, 624, and 816 CPU processors; the largest of which yields roughly 26 million degrees of freedom. In addition, we consider a relative tolerance $\varepsilon_\mathrm{rel}$ for the internal CG-AMG solvers of $10^{-8}$, $10^{-4}$, and $10^{-2}$. 

\begin{figure*}[p]
    \centering
    \def\svgwidth{0.75\textwidth}
    \input{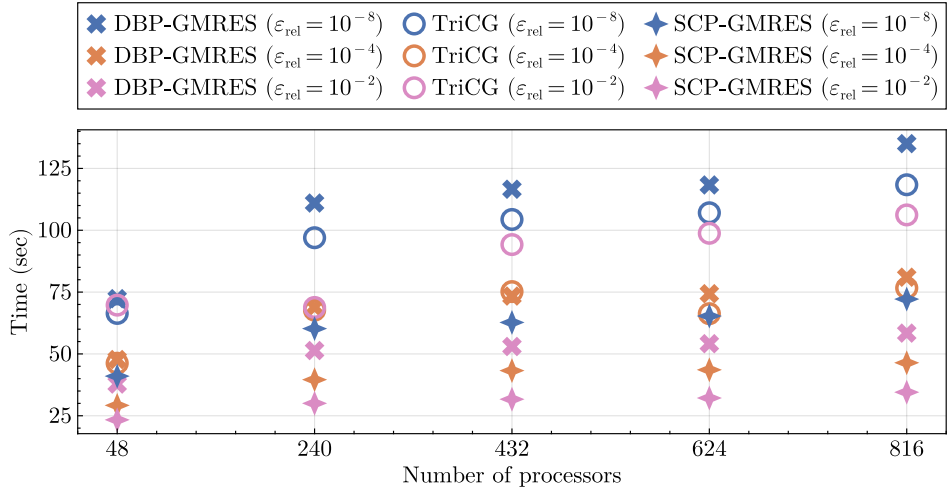}
    \caption{Weak scaling benchmarks for DBP-GMRES, TriCG, and SCP-GMRES where the number of degrees of freedom per processor is fixed at 32,000, and the number of processors is increased. $\varepsilon_\mathrm{rel}$ denotes the relative tolerance of the CG-AMG solvers for the elasticity and conductivity blocks.}
    \label{fig:weak_scaling_fig}
\end{figure*}
\begin{table*}[p]
    \small
    \centering
    \begin{tabular}{lllllll}
\hline 
Method &  & \multicolumn{5}{l}{Number of processors} \\\cline{3-7}
 &  & 48 & 240 & 432 & 624 & 816 \\
\hline 
DBP-GMRES & $\boldsymbol{v}\mapsto A\backslash \boldsymbol{v}$ Iters. & 34 & 51 & 56 & 52 & 59 \\
($\varepsilon_\mathrm{rel}=10^{-8}$) & $\boldsymbol{u}\mapsto -C\backslash \boldsymbol{u}$ Iters. & 17 & 20 & 18 & 20 & 20 \\
 & Global Iters. & 21 & 21 & 21 & 21 & 21 \\\cline{2-7}
DBP-GMRES & $\boldsymbol{v}\mapsto A\backslash \boldsymbol{v}$ Iters. & 17 & 26 & 28 & 27 & 30 \\
($\varepsilon_\mathrm{rel}=10^{-4}$) & $\boldsymbol{u}\mapsto -C\backslash \boldsymbol{u}$ Iters. & 8 & 8 & 8 & 9 & 9 \\
 & Global Iters. & 21 & 21 & 21 & 21 & 21 \\\cline{2-7}
DBP-GMRES & $\boldsymbol{v}\mapsto A\backslash \boldsymbol{v}$ Iters. & 8 & 12 & 13 & 13 & 14 \\
($\varepsilon_\mathrm{rel}=10^{-2}$) & $\boldsymbol{u}\mapsto -C\backslash \boldsymbol{u}$ Iters. & 4 & 4 & 4 & 4 & 4 \\
 & Global Iters. & 23 & 23 & 23 & 23 & 23 \\\cline{2-7}
TriCG & $\boldsymbol{v}\mapsto A\backslash \boldsymbol{v}$ Iters. & 31 & 42 & 46 & 45 & 52 \\
($\varepsilon_\mathrm{rel}=10^{-8}$) & $\boldsymbol{u}\mapsto -C\backslash \boldsymbol{u}$ Iters. & 17 & 18 & 19 & 19 & 20 \\
 & Global Iters. & 20 & 20 & 20 & 20 & 20 \\\cline{2-7}
TriCG & $\boldsymbol{v}\mapsto A\backslash \boldsymbol{v}$ Iters. & 12 & 18 & 19 & 20 & 21 \\
($\varepsilon_\mathrm{rel}=10^{-4}$) & $\boldsymbol{u}\mapsto -C\backslash \boldsymbol{u}$ Iters. & 8 & 10 & 9 & 9 & 10 \\
 & Global Iters. & 23 & 25 & 27 & 21 & 23 \\\cline{2-7}
TriCG & $\boldsymbol{v}\mapsto A\backslash \boldsymbol{v}$ Iters. & 7 & 8 & 11 & 11 & 12 \\
($\varepsilon_\mathrm{rel}=10^{-2}$) & $\boldsymbol{u}\mapsto -C\backslash \boldsymbol{u}$ Iters. & 4 & 4 & 4 & 4 & 6 \\
 & Global Iters. & 58 & 45 & 58 & 58 & 58 \\\cline{2-7}
SCP-GMRES & $\boldsymbol{v}\mapsto A\backslash \boldsymbol{v}$ Iters. & 34 & 51 & 56 & 52 & 59 \\
($\varepsilon_\mathrm{rel}=10^{-8}$) & $\boldsymbol{u}\mapsto \bar{S}\backslash \boldsymbol{u}$ Iters. & 16 & 18 & 17 & 18 & 19 \\
 & Global Iters. & 10 & 10 & 10 & 10 & 10 \\\cline{2-7}
SCP-GMRES & $\boldsymbol{v}\mapsto A\backslash \boldsymbol{v}$ Iters. & 17 & 26 & 28 & 27 & 30 \\
($\varepsilon_\mathrm{rel}=10^{-4}$) & $\boldsymbol{u}\mapsto \bar{S}\backslash \boldsymbol{u}$ Iters. & 7 & 9 & 8 & 9 & 9 \\
 & Global Iters. & 10 & 10 & 10 & 10 & 10 \\\cline{2-7}
SCP-GMRES & $\boldsymbol{v}\mapsto A\backslash \boldsymbol{v}$ Iters. & 8 & 12 & 13 & 13 & 14 \\
($\varepsilon_\mathrm{rel}=10^{-2}$) & $\boldsymbol{u}\mapsto \bar{S}\backslash \boldsymbol{u}$ Iters. & 4 & 4 & 4 & 4 & 4 \\
 & Global Iters. & 10 & 10 & 10 & 10 & 10 \\\hline 
\end{tabular}

    \caption{The iteration counts for each solver as the number of processors is increased. The first two rows of iteration counts denote the typical number of iterations to solve the corresponding maps using CG-AMG, while \textit{Global iters.} shows the number of iterations for the outer solver. We use a relative tolerance of $10^{-8}$ as the stopping criteria for the outer solvers.}
    \label{tab:iter_table}
\end{table*}

Figure \ref{fig:weak_scaling_fig} shows the weak scaling benchmarks for DBP-GMRES, TriCG, and SCP-GMRES. Table \ref{tab:iter_table} shows the iteration counts for the outer solver (GMRES/TriCG), inner CG-AMG solvers, and the final residual values for the outer solvers. The table shows that all three methods demonstrate weak scalability in terms of the global iteration counts. Surprisingly, TriCG yields only a small improvement compared to DBP-GMRES in terms of global iterations, however it does demonstrate a more significant improvement in terms of time to solve. In addition, the table shows that TriCG requires high quality solutions to the inner elasticity ($\boldsymbol{v}\mapsto A\backslash \boldsymbol{v}$) and conductivity ($\boldsymbol{u}\mapsto S\backslash \boldsymbol{u}$) blocks to ensure reasonable global iteration counts. SCP-GMRES demonstrates the best performance of the considered solvers owing to the efficient assembly of the approximate Schur complement and a rough halving of the global iteration count. The global iterations of the method also appear to be insensitive to the large relative tolerance of the CG-AMG solvers; this contributes to a significant reduction in overall time to solve.  

Although these solvers generally show an acceptable level of weak scalability, there is still room for improvement. For example, we see an increase in the iterations of the CG-AMG solvers as the number of degrees of freedom increases due to the AMG preconditioner not being weakly scalable. In future, additional focus should be given to the development of efficient and scalable solvers for problems involving the equations of piezoelectricity.

\subsection{Constrained optimisation and optimisation parameters}
In this paper we utilise the Hilbertian projection method developed by \cite{Wegert_2023b} to handle constrained optimisation in the level-set context. The Hilbertian projection method falls under the general class of null-space methods \citep{10.3934/dcdsb.2019249,10.1051/cocv/2020015_2020}. The method aims to improve the optimisation objective and constraints to first-order by choosing a normal velocity for the Hamilton-Jacobi evolution equation that is a combination of an orthogonal projection operator applied to the objective sensitivity, and a weighted sum of orthogonal basis functions for the constraint sensitivities. We utilise the distributed implementation of the Hilbertian projection method provided by \cite{GridapTopOpt}.

For all optimisation results in Section \ref{sec:Results and discussion}, we take the initial level-set function depicted by the solid phase in Figure \ref{fig:partitioned_domain}. The background computational mesh has $100^3$ first-order hexahedral elements and is distributed over 150 processors using (5,5,6)-partitioning. The mesh size could be increased above $100^3$ elements, however we choose to limit the computational resources used per optimisation to enable us to solve large number of topology optimisation problems. Material properties are smoothed within two elements of the zero level-set contour. We reinitialise the level-set function as a signed distance function at the beginning of the optimisation procedure and at every five subsequent iterations by solving the reinitialisation equation using a standard first order scheme. We refer the reader to \cite{GridapTopOpt} and the references therein for further details. In addition, for all optimisation problems described in the following section we use the Schur complement preconditioned GMRES method with inner solver and outer solver relative tolerances of $10^{-2}$ and $10^{-12}$, respectively. Finally, we impose a stopping criterion at an iteration $q\geq10$ that requires
\begin{equation}
    \frac{\lvert \mathcal{J}(\Omega^{q})-\mathcal{J}(\Omega^{q-j})\rvert}{\lvert \mathcal{J}(\Omega^{q})\rvert}< 0.2\epsilon_{m},~\forall j = 1,\dots,5,
\end{equation}
and
\begin{equation}
    \lvert C_i(\Omega^{q})\rvert < 0.001,~\forall i=1,\dots,N,
\end{equation}
where $\epsilon_{m}$ is the maximum side length of an element, $C_i$ is the value of the ith constraint, and there are $N$ constraints in total. 

It is important to note that inverse homogenisation problems, such as the ones considered in this paper, have non-unique solutions because translation of the optimised solid phase relative to the unit cell yields the same material properties. In addition, problems can have several local optima that are difficult for an optimiser to avoid. We checked initial structure dependence of these problems by solving the optimisation problems given in Equations \ref{eqn:optim prob I} and \eqref{eqn:optim prob 2} for several different initial structures. We found that the final objective values were similar across all considered initial structures.

\begin{figure*}[b]
    \centering
    \def\svgwidth{0.8\textwidth}
    \input{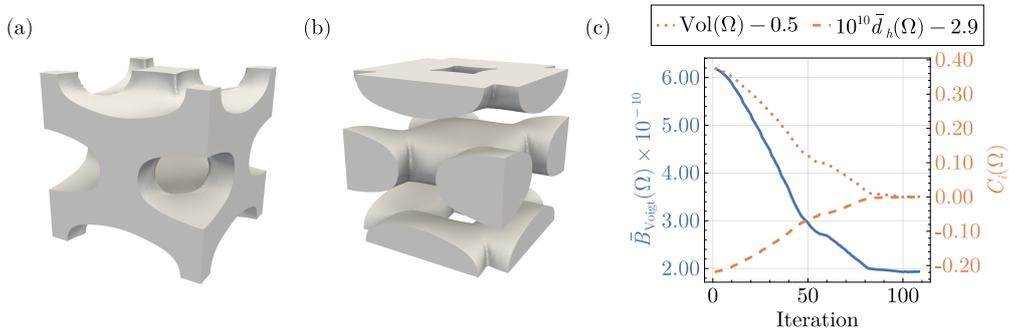}
    \caption{An example optimisation history for maximising $\bar{B}_\mathrm{Voigt}$ subject to constraints $\operatorname{Vol}(\Omega)=0.5$ and $\bar{d}_h(\Omega)=2.9\times10^{-10}$. Sub-figure~(a) shows the optimised solid phase ($\phi\leq0$), sub-figure~(b) shows the resulting void phase ($\phi\geq0$), and sub-figure~(c) shows the optimisation history. The iteration history of the objective value is given in blue, while the iteration history of the constraint values are given in orange.}
    \label{fig:BVoigt-sample-history}
\end{figure*}

\section{Results and discussion}
\label{sec:Results and discussion}

\subsection{Maximising Voigt/Reuss bulk modulus}\label{sec: max voigt/reuss}
In this section we consider the optimisation problem posed in Eq.~\eqref{eqn:optim prob I} for maximising the effective Voigt or Reuss bulk modulus subject to a 50\% volume constraint and a constraint on the effective hydrostatic strain coefficient. Figure \ref{fig:BVoigt-sample-history} shows an example optimised unit-cell and iteration history for maximising $\bar{B}_\mathrm{Voigt}$ subject to constraints $\operatorname{Vol}(\Omega)=0.5$ and $\bar{d}_h(\Omega)=2.9\times10^{-10}$. All optimised unit-cells are shown with $\phi\leq0$ for the solid phase and $\phi\geq0$ for the void phase. The iteration history (Figure~\ref{fig:BVoigt-sample-history}c) shows that the optimiser converges after roughly 180 iterations and the objective and constraint histories are fairly smooth even under topological changes. As expected, we do not observe large regions of intermediate densities owing to use of the level-set method.

\begin{figure*}[p]
    \centering
    \def\svgwidth{0.70\textwidth}
    \input{Figure4_svg-tex.tex}
    \caption{Optimisation results and their properties (blue circles) obtained by maximising $\bar{B}_\mathrm{Voigt}$ (x-axis) subject to $\operatorname{Vol}(\Omega)=0.5$ and a prescribed  $\bar{d}_h(\Omega)$ (y-axis). The orange crosses show the resulting Voigt bulk modulus from maximising $\bar{B}_\mathrm{Reuss}$, the structures of which are shown in Figure \ref{fig:BReuss-dh}. We highlight one structure and its properties for further discussion in Section \ref{sec: max voigt/reuss}.}
    \label{fig:BVoigt-dh}
    \centering
    \def\svgwidth{0.70\textwidth}
    \input{Figure5_svg-tex.tex}
    \caption{Optimisation results and their properties (orange crosses) obtained by maximising $\bar{B}_\mathrm{Reuss}$ (x-axis) subject to $\operatorname{Vol}(\Omega)=0.5$ and a prescribed  $\bar{d}_h(\Omega)$ (y-axis). The blue circles show the resulting Reuss bulk modulus from maximising $\bar{B}_\mathrm{Voigt}$, the structures of which are shown in Figure \ref{fig:BVoigt-dh}. We highlight one structure and its properties for further discussion in Section \ref{sec: max voigt/reuss}. The microstructures marked by $\vardiamondsuit$, $\varheartsuit$, and $\spadesuit$ are considered in Table~\ref{tab:cutfem-verif}.}
    \label{fig:BReuss-dh}
\end{figure*}

Figures \ref{fig:BVoigt-dh} and \ref{fig:BReuss-dh} show the resulting effective Voigt and Reuss bulk modulus, respectively, for unit cells that are optimised for either effective measure of bulk modulus. We plot these results for different values of required effective hydrostatic strain coefficient and show the resulting solid phase of the unit cells.
\begin{figure*}[!t]
    \centering
    \def\svgwidth{0.7\textwidth}
    \input{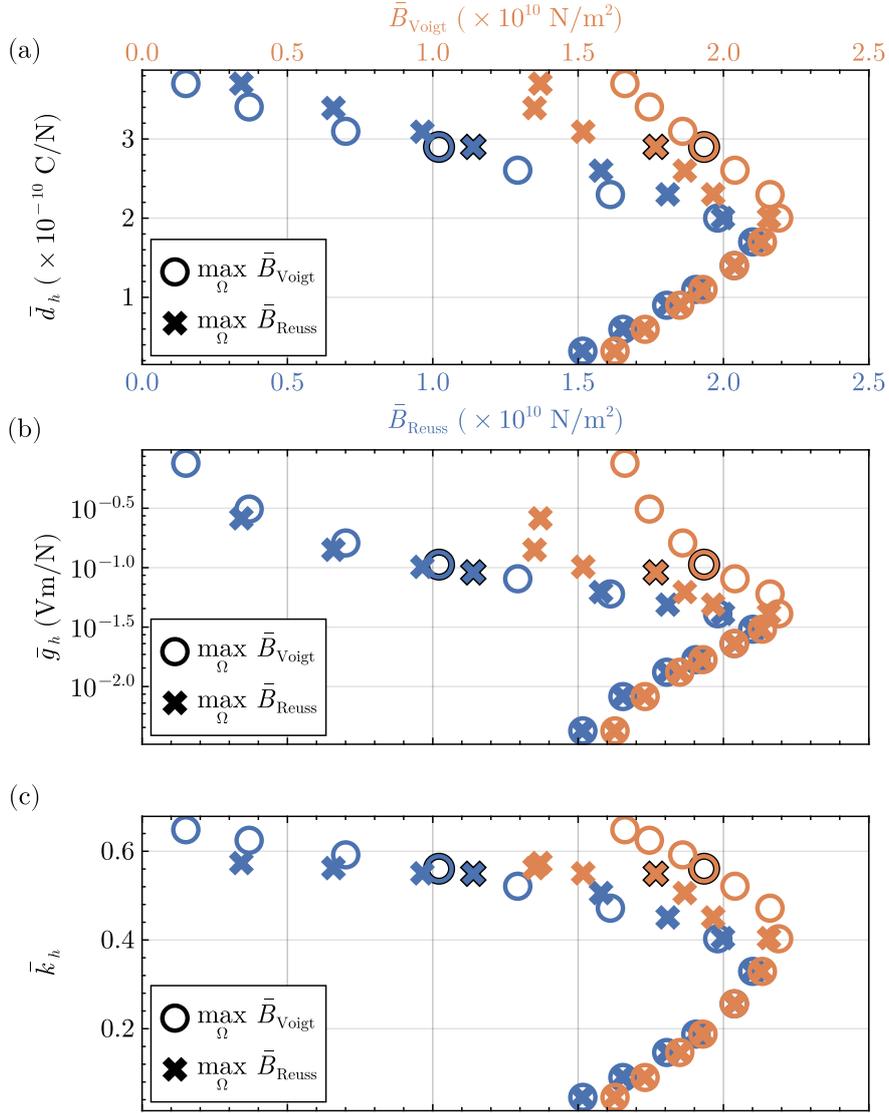}
    \caption{Visualisation of $\bar{B}_{\mathrm{Voigt}}$ (top x-axis) and $\bar{B}_{\mathrm{Reuss}}$ (bottom x-axis) for the optimisation results in Figures \ref{fig:BVoigt-dh} and \ref{fig:BReuss-dh} for maximising the Voigt (circles) or Reuss (crosses) bulk modulus. The orange and blue markers show the resulting Voigt and Reuss bulk modulus, respectively. Sub-figures (a), (b), and (c) show the effective hydrostatic strain coefficient, effective voltage constant, and the effective electromechanical coupling factor on the y-axis, respectively. The x-axis limits are the same for all sub-figures. The markers with black outlines correspond to the properties of the highlighted structures in Figures \ref{fig:BVoigt-dh} and \ref{fig:BReuss-dh}.}
    \label{fig:Bboth-foms}
\end{figure*}
Figure \ref{fig:Bboth-foms} visualises the effective hydrostatic strain coefficient, effective voltage constant, and effective electromechanical coupling, along with the corresponding values of Voigt and Reuss bulk modulus, for the optimised unit cells in Figures \ref{fig:BVoigt-dh} and \ref{fig:BReuss-dh}.

The results in both Figures \ref{fig:BVoigt-dh} and \ref{fig:BReuss-dh} show a clean transition between optimised microstructures for different values of the constraint on $\bar{d}_h$. The largest Voigt and Reuss bulk modulus is achieved with a closed microstructure at the right-most data point in these figures. This is expected because open-cell structures produce a sub-optimal bulk modulus \citep{Sigmund_1999}. Increasing the constraint on $\bar{d}_h$ from the aforementioned right-hand point reduces the corresponding optimised bulk modulus measure. In the case of microstructures optimised for $\bar{B}_\mathrm{Voigt}$, increasing $\bar{d}_h$ produces thin members connecting thick base-plates (see Figure~\ref{fig:BVoigt-dh}). These structures have a small effective stiffness $\bar{C}_{zzzz}$ in the poling direction for large $\bar{d}_h$. They also show a significant reduction of the Reuss bulk modulus when compared to the Voigt bulk modulus (compare the blue circles in Figure~\ref{fig:BVoigt-dh} and \ref{fig:BReuss-dh}, respectively). These features appear because the Voigt bulk modulus $\bar{B}_\mathrm{Voigt}$ is computed as an average of components of the stiffness tensor while the Reuss bulk modulus is computed as the reciprocal of the hydrostatic compliance \citep{Hill_1952}. As a result, a small $\bar{C}_{zzzz}$ produces a minor reduction in the Voigt bulk modulus but a significant reduction for the Reuss bulk modulus. This phenomenon is clearly shown by the distance between the circles in Figure \ref{fig:Bboth-foms}a. 

Suppose we now consider maximising the Reuss bulk modulus for large values of the constraint on $\bar{d}_h$ (see top left of Figure~\ref{fig:BReuss-dh}). When compared to the results in the upper left corner of Figure \ref{fig:BVoigt-dh}, we see that this objective produces more complex geometric features with robust members in the poling direction. This is particularly advantageous as these optimised microstructures are much stiffer in the poling direction than their maximum Voigt bulk modulus counterparts while also producing an enhanced hydrostatic strain coefficient. To the knowledge of the authors, this is the largest obtained hydrostatic strain coefficient value in the three-dimensional piezoelectric microstructure optimisation literature.

\begin{table}[]
\centering
\renewcommand{\arraystretch}{1.3}
\begin{tabular}{c|lll|lll}
             & \multicolumn{3}{c|}{Boundary smoothing} & \multicolumn{3}{c}{CutFEM}     \\ \cline{2-7} 
 Marker &
  \multicolumn{1}{l}{$\bar{d}_h$ (C/N)} &
  \multicolumn{1}{l}{$\bar{B}_\mathrm{Reuss}$ (Pa)} &
  \multicolumn{1}{l|}{$\bar{B}_\mathrm{Voigt}$ (Pa)} &
  \multicolumn{1}{l}{$\bar{d}_h$ (C/N)} &
  \multicolumn{1}{l}{$\bar{B}_\mathrm{Reuss}$ (Pa)} &
  \multicolumn{1}{l}{$\bar{B}_\mathrm{Voigt}$ (Pa)} \\ \hline
$\vardiamondsuit$  & $3.20\times10^{-11}$    & $1.51\times10^{10}$    & $1.63\times10^{10}$    & $3.20\times10^{-11}$ & $1.50\times10^{10}$ & $1.62\times10^{10}$ \\ \hline
$\varheartsuit$ & $1.70\times10^{-10}$    & $2.10\times10^{10}$    & $2.13\times10^{10}$    & $1.70\times10^{-10}$ & $2.09\times10^{10}$ & $2.12\times10^{10}$ \\ \hline
$\spadesuit$     & $3.70\times10^{-10}$    & $3.42\times10^{9}$    & $1.37\times10^{10}$    & $3.71\times10^{-10}$ & $3.05\times10^{9}$ & $1.33\times10^{10}$
\end{tabular}
\caption[Comparison of material properties computed using the boundary smoothing approach and CutFEM]{Comparison of the effective material properties computed using the boundary smoothing and CutFEM approaches for the three microstructures in Figure~\ref{fig:BReuss-dh} marked by $\vardiamondsuit$, $\varheartsuit$, and $\spadesuit$.}
\label{tab:cutfem-verif}
\end{table}
The smoothing of material properties near the zero level-set contour introduces error into the finite element approximation. In Table~\ref{tab:cutfem-verif} we present the effective material properties of three of the optimised microstructures in Figure~\ref{fig:BReuss-dh} computed with both the boundary smoothing approach and the more accurate CutFEM unfitted finite element approach \citep{10.1002/nme.4823_2015}. Overall, the small differences between the CutFEM and boundary smoothing values in Table~\ref{tab:cutfem-verif}  show that the boundary smoothing approach appropriately captures the effective material properties of interest. However, there is approximately a 12\% error when calculating the Reuss bulk modulus with the smoothed boundary approach for the top-left microstructure in Figure~\ref{fig:BReuss-dh} (marked $\spadesuit$). Such a discrepancy arises due to the fine-scale features present in the microstructure. This highlights the need for future research into more accurate techniques for capturing the material boundary in level-set based topology optimisation to find robust microstructures with more extreme hydrostatic strain coefficients. Similarly, further increase of the hydrostatic strain coefficient in problems optimised for the Voigt bulk modulus requires either a higher mesh resolution to account for fine-scale features or more advanced boundary capturing methods. Minimum length-scale constraints could alternatively be applied to avoid fine-scale features \citep[e.g.,][]{10.1007/s00158-016-1453-y_2016}, however the construction and implementation of these is currently non-trivial in the context of level-set methods. The need for the investigation and use of the aforementioned methods in future is also evidenced by the optimised microstructure in the top left of Figure \ref{fig:BVoigt-dh}, in which joints at the top and bottom of the microstructure are connected by weak material owing to the diffused boundary. In future, we plan to address these issues using unfitted finite elements for level set-based topology optimisation as proposed in \citet{Wegert_Manyer_Mallon_Badia_Challis_2025}.

The results shown in Figure \ref{fig:BVoigt-dh} improve upon those shown in \cite{Wegert_2022} and provide additional insights. In particular, the results for maximising the Voigt bulk modulus for large $\bar{d}_h$ give higher $\bar{B}_\mathrm{Voigt}$ than those in \cite{Wegert_2022}. In addition, we are able to capture more information regarding the $\bar{B}_\mathrm{Voigt}$--$\bar{d}_h$ parameter space as we avoid the use of a weighted sum of objectives. Specifically, our results here suggest that the cross-property space is not convex as was previously hypothesised. That is, for $\bar{d}_h<2\times10^{-10}$ (in the case of Fig. \ref{fig:BVoigt-dh}) the optimised bulk modulus reduces as the hydrostatic strain coefficient constraint is decreased. In this region of the parameter space the optimised microstructures tend to become extruded out from the $xy$-plane as $\bar{d}_h$ approaches zero. We believe that this occurs because the $\bar{d}_{zxx}$ and $\bar{d}_{zyy}$ components of the piezoelectric coupling tensor are becoming large to cancel part of $\bar{d}_{zzz}$ under hydrostatic loading. The results in the lower band of Figures \ref{fig:BVoigt-dh} and \ref{fig:BReuss-dh} also show that the resulting Voigt bulk modulus or Reuss bulk modulus are almost identical when optimising either measure of bulk modulus for sufficiently small hydrostatic strain coefficient. We hypothesise that this is a result of having a lower measure of anisotropy for these structures, while structures in the upper band of the optimised results with $\bar{d}_h>1.7\times10^{-10}$ present increased anisotropy owing to the reduced stiffness in the poling direction.

The effective piezoelectric figures of merit for these optimised materials as shown in Figure \ref{fig:Bboth-foms} demonstrate the capability of the optimised periodic materials for engineering applications. Materials that are optimised for the Voigt bulk modulus generally demonstrate the largest values for the effective voltage constant $\bar{g}_h$ and electromechanical coupling figures of merit $\bar{k}_h$ owing to their reduced stiffness in the poling direction. However, the thin members that are present in the microstructures make them less desirable from a manufacturing point of view than the microstructures optimised for the Reuss bulk modulus. 

From an engineering perspective, structures at extreme values of either the effective hydrostatic strain coefficient or effective bulk modulus may not be manufacturable without significant difficulty and cost. For extreme bulk modulus values, closed-cell structures are present that are difficult to manufacture. On the other hand, thin members and complex geometric features are present for extreme values of the hydrostatic strain coefficient. It is therefore important to highlight structures that present excellent properties with no fine-scale features. In Figures \ref{fig:BVoigt-dh} and \ref{fig:BReuss-dh} we highlight two such structures that correspond to $\bar{d}_h=2.9\times10^{-10}$. These open-cell structures exhibit no fine-scale features while possessing excellent properties. In particular, these structures provide a ${\sim}9$ times increase of the base material $d_h$; a ${\sim}43$ and ${\sim}50$ times increase of the base material $g_h$, respectively; and a ${\sim}7$ times increase of the base material $k_h$. These properties are achieved with a 50\% volume fraction at roughly 20\% and 11\% of the base material $B_\mathrm{Voigt}$ and $B_{\mathrm{Reuss}}$, respectively. Of course, different materials in these families of optimised structures may be more appropriate for specific applications. In future, we plan to consider additive manufacturing constraints in the optimisation problem for metamaterials with extreme piezoelectric properties.

\subsection{Maximising hydrostatic strain coefficient}
In this section we consider the optimisation problem posed in Eq.~\eqref{eqn:optim prob 2} to maximise the effective hydrostatic strain coefficient $\bar{d}_h$ subject to a constraint on the effective stiffness $\bar{C}_{zzzz}$ in the poling direction. Figure \ref{fig:max-dh-figs} and Table \ref{tab:max_dh_results} show the resulting optimisation results and the figures of merit for $\bar{C}_{zzzz}(\Omega)=5\times10^9$, $2.5\times10^9$, and $1\times10^9$ (Pa).
\begin{figure*}[!t]
    \centering
    \def\svgwidth{\linewidth}
    \input{Figure7_svg-tex.tex}
    \caption{Optimised solid phase ($\phi\leq0$) that maximises $\bar{d}_h$ subject to $\bar{C}_{zzzz}(\Omega)=5\times10^9$, $2.5\times10^9$, and $1\times10^9$ shown in sub-figures~(a), (b), and (c), respectively. In sub-figure~(a) we also show the iteration history for the corresponding problem.}
    \label{fig:max-dh-figs}
    \begin{tabular}{l|lllllll}
Fig. & Required $\bar{C}_{zzzz}$ (Pa) & $\max\bar{d}_h$ (C/N) & $\mathrm{Vol}$ & $\bar{B}_\mathrm{Voigt}$ (Pa) & $\bar{B}_\mathrm{Reuss}$ (Pa) & $\bar{g}_h$ (Vm/N) & $\bar{k}_h$ \\
\hline 
(a) & 5.0$\times10^{9}$ & 3.5$\times10^{-10}$ & 0.599 & 1.75$\times10^{10}$ & 4.85$\times10^{9}$ & 0.197 & 0.578 \\
\hline 
(b) & 2.5$\times10^{9}$ & 3.89$\times10^{-10}$ & 0.586 & 1.63$\times10^{10}$ & 2.45$\times10^{9}$ & 0.357 & 0.584 \\
\hline 
(c) & 1.0$\times10^{9}$ & 4.45$\times10^{-10}$ & 0.581 & 1.58$\times10^{10}$ & 9.77$\times10^{8}$ & 0.889 & 0.622 \\
\hline 
\end{tabular}

    \captionof{table}{Resulting properties and piezoelectric figures of merit for the optimised microstructures in Figure \ref{fig:max-dh-figs}.}
    \label{tab:max_dh_results}
\end{figure*}

The results in Figure \ref{fig:max-dh-figs} and Table \ref{tab:max_dh_results} demonstrate that large piezoelectric figures of merit can be obtained by maximising the hydrostatic strain coefficient with only a constraint on the stiffness in the poling direction. The resulting structures in Figure \ref{fig:max-dh-figs} vary most near their top and bottom faces. This is because we only consider relatively small changes to the required stiffness $\bar{C}_{zzzz}$ and the optimiser achieves this by thinning the geometric features near the top and bottom faces. This suggests that the effective hydrostatic coupling is quite sensitive to high compliance. The resulting geometries in Figure \ref{fig:max-dh-figs} also present additional complex geometric features that may be difficult to manufacture in practice. This makes these types of problems ideal candidates for adding manufacturing constraints in future work.

We note that similar to the previous optimisation problem, further reduction of the stiffness $\bar{C}_{zzzz}$ produces joints that are connected by weak material due to the diffused boundary (see Figure~\ref{fig:max-dh-figs}c). As discussed above in Section~\ref{sec: max voigt/reuss}, we plan to address this issue in future work by using unfitted finite elements with level-set topology optimisation.

\section{Conclusions}
\label{sec: Conclusions}
In this paper we have presented results for level set-based topology optimisation of periodic piezoelectric microstructures. To make these problems possible at high computational resolutions, we use memory-distributed computing techniques. We implement this using GridapTopOpt \citep{GridapTopOpt}, an extendable framework for domain partitioned PDE-constrained optimisation. We emphasise that serial implementations would be severely time and memory-bound. We considered four block-preconditioned linear solvers and found that the approximate Schur complement preconditioned GMRES method (SCP-GMRES) demonstrates the best performance and scalability owing to the efficient assembly of the approximate Schur complement, a rough halving of the global iteration count, and insensitivity to large relative tolerances of inner solvers. In future we plan to further develop efficient and scalable solvers for large-scale piezoelectric topology optimisation.

In this work we considered two families of optimisation problems. The first maximises the effective Voigt or Reuss bulk modulus subject to a constraint on the volume and the effective hydrostatic strain coefficient, while the second maximises the effective hydrostatic strain coefficient subject to a constraint on the effective stiffness in the poling direction. In the results for both of these we observed that, as expected, the large regions of intermediate densities that previously increased the optimisation objective non-physically when using density-based methods were remedied via use of the level-set method \citep[c.f.,][]{10.1016/j.jcp.2003.09.032_2004}. We also found that a finer computational resolution in both problems allowed us to obtain enhanced figures of merit, well above previous results in the piezoelectric material optimisation literature. 

For bulk modulus maximisation problems, we found that optimising for the Reuss measure of bulk modulus instead of the Voigt measure produces more robust microstructures with thicker members in the poling direction. This is particularly advantageous because microstructures that maximise the Reuss bulk modulus are much stiffer in the poling direction than their counterparts that maximise the Voigt bulk modulus, while also producing an enhanced hydrostatic strain coefficient. In addition, we are able to capture more information regarding the $\bar{B}_\mathrm{Voigt}$--$\bar{d}_h$ parameter space compared to earlier work because we reformulate the optimisation problem posed in \cite{Wegert_2022} to avoid the linear combination of $\bar{B}_\mathrm{Voigt}$ and $\bar{d}_h$. Our new results suggest that the parameter space $\bar{B}_\mathrm{Voigt}$--$\bar{d}_h$ is not convex, in contrast to what was previously hypothesised in \cite{Wegert_2022}. In addition, the results for smaller values of the constraint on $\bar{d}_h$ show that the effective Voigt or Reuss bulk modulus, respectively, are almost identical when optimising either measure of bulk modulus for a sufficiently small hydrostatic strain coefficient. We hypothesise that this is a result of the corresponding structures having a lower measure of anisotropy, whereas the upper band presents increased anisotropy due to the reduction of stiffness in the poling direction.

We found that optimised materials that are either constrained to have large hydrostatic strain coefficient $\bar{d}_h$ or that maximise $\bar{d}_h$ possess enhanced effective figures of merit for engineering applications such as sensor, hydrophone, or actuator design. We found that materials that are optimised for the Voigt bulk modulus generally demonstrate the largest values for the effective voltage constant $\bar{g}_h$ and electromechanical coupling figures of merit $\bar{k}_h$ owing to their reduced stiffness in the poling direction. However, the thin members present in the microstructures make them less desirable from a manufacturing point of view compared to the microstructures optimised for the Reuss bulk modulus. 

In our results and discussion, we highlighted two optimised materials from Figure \ref{fig:BVoigt-dh} and \ref{fig:BReuss-dh} that offer a reasonable trade-off between manufacturability, mechanical properties, and piezoelectric figures of merit. The highlighted optimised structures possess no fine-scale features, while exhibiting excellent piezoelectric figures of merit that are several times larger than those of the base material.

To conclude, we note potential directions for future research. The inclusion of manufacturing constraints as part of these optimisation problems would help improve the manufacturability of the proposed optimised metamaterials, particularly for metamaterials possessing extreme piezoelectric properties. Furthermore, the future use of unfitted finite elements for level set-based topology optimisation would allow for more accurate capture of the material boundary and enable the discovery of microstructures with more extreme hydrostatic strain coefficients. Finally, multi-material piezoelectric topology optimisation would enable the computational design and optimisation of more advanced metamaterials including multi-poled piezoelectric materials or piezoelectric materials that are embedded in other types of materials such as soft materials.  

\section*{CRediT authorship contribution statement}
\textbf{Zachary J Wegert:} Writing -- original draft, Writing -- review and editing, Conceptualization, Data curation, Formal analysis, Investigation, Methodology, Software, Validation, Visualization. \textbf{Anthony P Roberts:} Writing -- review and editing, Conceptualization, Funding acquisition, Supervision. \textbf{Vivien J Challis:} Writing -- review and editing, Conceptualization, Funding acquisition, Project administration, Supervision.

\section*{Declaration of Competing Interest}
The authors have no competing interests to declare that are relevant to the content of this article.

\section*{Data availability}
The source code and data for this work is available at \url{https://github.com/zjwegert/Wegert_et_al_2024b}.

\section*{Acknowledgement}
This work was supported by the Australian Research Council through the Discovery Projects grant scheme (DP220102759). This research used computational resources provided by: the eResearch Office, Queensland University of Technology; the Queensland Cyber Infrastructure Foundation (QCIF); The University of Queensland's Research Computing Centre (RCC); and the National Computational Infrastructure (NCI) Australia. The first author is supported by a QUT Postgraduate Research Award and a Supervisor Top-Up Scholarship. The above support is gratefully acknowledged. The first author would like to thank Jordi Manyer for several useful discussions regarding numerical solvers.

\appendix


 \bibliography{main}





\end{document}